\documentclass[review]{elsarticle}

\usepackage[utf8]{inputenc}
\usepackage{lineno,hyperref}
\modulolinenumbers[5]
\usepackage{amsmath}
\usepackage{amsfonts}
\usepackage{graphicx}

\journal{Journal of \LaTeX\ Templates}

\begin{document}

\begin{frontmatter}

\title{Implication of the Superstring theory on Astrophysics and Cosmology}

\author{Hongsu Kim\fnref{myfootnote}}
\address{Center for Theoretical Astronomy, Korea Astronomy and Space Science Institute, Daejeon 34055, Republic of Korea}
\fntext[myfootnote]{chris@kasi.re.kr}

\begin{abstract}
Under the spirit of J. Polchinsky's introduction of D-brane in the string target spacetime (or 'bulk', for short), we consider the presence of both closed and open superstrings and coupling of their low-lying modes in the bulk. As a probe to explore the astrophysical and cosmological implications of the superstring theory at larger length scale or in low-energy dynamics, we construct the Yang-Mills monopole black hole solution and investigate its features. Our results exhibit that as a result of intimate interactions among low-lying modes, some cherished features of classical gravitation and cosmology break down. They are the violation of black hole no-hair theorem and the break down of cosmic censorship hypothesis. We learn from this lesson that these implications of fully developed superstring theory in classical gravitation and cosmology largely depart from our conventional wisdom signaling that the most current version of superstring theory is indeed the best candidate for quantum theory of gravitation.
\end{abstract}

\begin{keyword}
Open and closed string, D-brane, Black Hole no-hair theorem, Cosmic censorship hypothesis
\end{keyword}

\end{frontmatter}

 \section{intoduction}
 Back in 1994, the Seiberg-Witten conjecture starting the discovery of string S-duality was followed by the discoveries of other string dualities like T-duality and Kaluza-Klein type 10 to 4 dimensional reduction. This avenue of development, later, has been dubbed, String Duality. Nearly in parallel, J. Polchinski's introduction of D-brane and J. Maldacena's associated proof of AdS/CFT duality which rapidly evolved into more comprehensive, Gravity/Gauge duality, later, have been coined, brane physics.
 By now, these two avenues of developments have joined together and collectively have been identified with the 2nd string revolution.
 
 Interestingly enough, although not being along the 2nd avenue of development, namely, brane physics, the author of the present article, Dr. Hongsu Kim has published as early as in 1996, an article in Phys. Letter B \cite{1} that we now would like to revisit and enlarge in a much more rich fashion.
 To get right onto the point, in the present work, we would like to exhibit that the above-mentioned 1996 article of the present author was indeed the first unidentified effort, where, in a manifest fashion, the generic features of brane physics and more comprehensively, the 2nd string revolution, namely, the manifest spacetime quantum fluctuation have actually been demonstrated in terms of, first the violation of the black hole no-hair theorem, and second the breakdown of cosmic censorship hypothesis.
 
 This is truly remarkable, as it implies that in the context of a legitimate quantum theory of gravitation, which, for now, is the string duality and brane physics or collectively, the full version of superstring theory, the longstanding, cherished and hence familiar features of classical theory of gravitation and cosmology manifestly break down and thus are not valid any longer.
 To the best of our knowledge, this type of effort toward the demonstration of the breakdown of cherished features of classical gravitation and cosmology has not appeared in the published literature, yet.
 
\section{The bulk domain where the closed and the open string low-lying modes couple.}
 By now, upon the advent of 2nd string revolution, and particularly in the context of the Brane Physics, all the 4 interactions of Nature, namely gauge theory and gravity are successfully unified and along this line obviously the major role is played by closed string perturbations in the bulk, and open string perturbations on the brane, that is, gravitation and gauge theory, respectively.\\
 To be more precise, as a consequence of closed string perturbations in the bulk particularly, its low-lying modes ($g_{\mu\nu}, B_{\mu\nu}, \phi$) and as a consequence of open string perturbations on the brane, particularly, its low-lying mode ($A^a_\mu$ \& $ F^a_\mu\nu$) any event that happens should be treated basically by starting with the Lagrangian that the present author dubbed, Einstein-Antisymmetric Tensor Theory in his earlier 1996 article \cite{1}. It is given by\\
 \begin{equation}
L = \frac{1}{8}\epsilon^{\mu\nu\rho\sigma}B^a_{\mu\nu}F^a_{\rho\sigma}-\frac{1}{8}A^a_\mu A^{a\mu}
 \end{equation}
 where the antisymmetric tensor gauge field $B^a_{\mu\nu}$ \cite{2, 3, 4, 5} and a vector potential $A^a_\mu$ associated with its field strength $F^a_{\mu\nu}$ are treated as being independent variables of the theory. Now by varying this Lagrangian with respect to these independent variables, we can obtain the classical field equation for $B^a_{\mu\nu}$ and $A^a_\mu$ respectively as 
\begin{align}
&F^a_{\mu\nu} = 0,\\
&D^\mu \tilde{B}^a_{\mu\nu} + A^a_\nu = 0 
\end{align}
where 
\begin{align}
&F^a_{\mu\nu} = \partial_\mu A^a_\nu - \partial_\nu A^a_\mu + g_cf^{abc}A^b_\mu A^c_\nu,\nonumber\\
&D^{ac}_\mu = (\partial_\mu\delta^{ac} + g_cf^{abc}A^b_\mu),\quad \tilde{B}^a_{\mu\nu} = \epsilon_{\mu\nu}^{\rho\sigma}B^a_{\rho\sigma}. \nonumber
\end{align}
Further, by acting the operator $D^\nu$ on the field equation for $A^a_\mu$ in eq.(3) and using eq.(2), we get
\begin{equation}
D^\mu A^a_\mu = 0.
\end{equation}
Then the vector potential $A^a_\mu$ satisfying classical field equation eqs.(2) and (3) turns out to be the "pure gauge",
\begin{equation}
A_\mu = -\frac{i}{g_c}(\partial_\mu U(x))U^{-1}(x)
\end{equation}
(where $U(x)$ is the gauge transformation function of the given non-abelian gauge group) provided it (eq.(5)) satisfies the "combined" field equation (4). Using the classical field equation for $A^a_\mu$ in eq.(3) one can show that the first-order formulation Lagrangian in eq.(1) turns into the second-order formulation Lagrangian \cite{6}. Further, since the classical field equation for $B^a_{\mu\nu}$ is the vanishing $F^a_{\mu\nu}$, in this first-order formulation $B^a_{\mu\nu}$ field appears classically as a Lagrange multiplier enforcing the constraint $F^a_{\mu\nu} = 0$. And finally, substituting the pure gauge solution in eq.(5) into the Lagrangian (1) demonstrates the equivalence of the theory to the non-linear $\sigma$-model \cite{6}. Now we consider the case when the gravity is turned on. To begin, it seems essential for us to declare our sign convention. We choose to take the convention in which $g_{\mu\nu} = \textrm{diag}(- + + +) $ and $R^a_{\beta\mu\nu} = \partial_\mu \Gamma^\alpha_{\beta\nu} - \partial_\nu \Gamma^\alpha_{\beta\mu} + \Gamma^\alpha_{\mu\lambda}\Gamma^\lambda_{\beta\nu}  - \Gamma^\alpha_{\nu\lambda}\Gamma^\lambda_{\beta\mu}$. It is crucial to fix the right sign for the antisymmetric tensor (i.e., the matter) sector of the action "relative" to the gravity action. Therefore in our sign convention, we explain the way we determined the sign for the matter action using the fact that on-shell, the antisymmetric tensor theory action is equivalent to that of non-linear sigma model with the right sign. Consider generators $T^a$ of the non-abelian group G in a representation in which $[T^a, T^b] = if^{abc}T^c,$ Tr$(T^aT^b) = c\delta^{ab}$ and $U(x)$ = exp$[i\phi^a(x)T^a]$ where $f^{abc}$ and $c$ are the structure constant and a representation dependent positive constant respectively. Then it can be readily shown that upon substituting the on-shell condition, $F_{\mu\nu} = 0$, i.e., $A_\mu = - \frac{i}{g_c}(\partial_\mu U)U^{-1}$, one gets $L = (\frac{1}{8c})$Tr$[\epsilon^{\mu\nu\rho\sigma}B_{\mu\nu}F_{\rho\sigma} - A_\mu A^\mu] = (-\frac{1}{8cg^2_c})$Tr$[(\partial_\mu U^{-1})(\partial^\mu U)]$ (with $B_{\mu\nu} = B^a_{\mu\nu}T^a$ and $A_\mu = A^a_\mu T^a)$ which is of the right sign. Thus in this way we have fixed the sign for the matter action. Now in order to describe the coupled Einstein antisymmetric tensor theory we again employ the first order formulation of the antisymmetric tensor sector, then the theory is described by the action  (we work in the unit G = 1)

\begin{align}
S &= S_G + S_{AT} \nonumber\\ 
&=\int d^4x\sqrt{g}\left[\frac{1}{16\pi}R + \frac{1}{8}\left(\frac{1}{\sqrt{g}}g^{\mu\alpha}g^{\nu\beta}\tilde{B}^a_{\mu\nu}F^a_{\alpha\beta} - g^{\mu\nu}A^a_\mu A^a_\nu\right)\right]
\end{align}
where we used that in curved spacetime, $\epsilon^{\mu\nu\rho\sigma} \rightarrow (\frac{\epsilon^{\mu\nu\rho\sigma}}{\sqrt{g}}) $ and again $\tilde{B}^a_{\mu\nu} = \epsilon_{\mu\nu}^{\rho\sigma}B^a_{\rho\sigma} = g_{\mu\alpha}g_{\nu\beta}\epsilon^{\alpha\beta\rho\sigma}B^a_{\rho\sigma}.$ The curved spacetime version of the classical field equation for $B^a_{\mu\nu}$ and $A^a_\mu$ are given respectively by
\begin{align}
&F^a_{\mu\nu} = 0,\\
&D^\mu\tilde{B}^a_{\mu\nu} + \sqrt{g}A^a_\nu = 0
\end{align}
along with the curved spacetime counterpart of eq.(4) which is the necessary condition that $A^a_\mu$ must satisfy as classical solution being given by
\begin{equation}
D^\mu(\sqrt{g}A^a_\mu) = 0.
\end{equation}
In addition, varying the action in eq.(6) with respect to the metric $g_{\mu\nu}$ yields the Einstein field equation $R_{\mu\nu}-\frac{1}{2}g_{\mu\nu}R=8\pi GT_{\mu\nu}$ where 
\begin{equation}
R_{\mu\nu}  = -4\pi\left[\frac{1}{\sqrt{g}} \{g^{\alpha\beta}(\tilde{B}^a_{\mu\alpha}F^a_{\nu\beta})-\frac{1}{2}g_{\mu\nu}(\tilde{B}^a_{\alpha\beta}F^{a\alpha\beta})\} - \frac{1}{2}(A^a_\mu A^a_\nu)\right]
\end{equation}
with the energy-momentum tensor being given by
\begin{equation*}
T_{\mu\nu} = -\frac{1}{8}\left[\frac{1}{\sqrt{g}}g^{\alpha\beta}(4\tilde{B}^a_{\mu\alpha}F^a_{\nu\beta}) + \{g_{\mu\nu}(A^a_\alpha A^{a\alpha} - 2(A^a_\mu A^a_\nu)\}\right]. 
\end{equation*} 
Our strategy for solving the classical equations of motion in eqs.(7), (8) and (10) along with the necessary condition eq.(9) is as follows; we start with the solution to the field equations (7) and (9) which, as we shall see, still turns out to be the pure gauge in eq.(5) even in the curved spacetime. Next, for this pure gauge solution satisfying eqs.(7) and (9), the Einstein field equation in (10) takes a remarkably simple form,  $R_{\mu\nu} = 2\pi(A^a_\mu A^a_\nu)$ with $T_{\mu\nu} = \frac{1}{4}[A^a_\mu A^a_\nu - \frac{1}{2}g_{\mu\nu}(A^a_\alpha A^{a\alpha})]$ that can be readily solvable. Finally by substituting the pure gauge solution for $A^a_\nu$ and the metric solution $g_{\mu\nu}$ into the field equation (8) (plus possibly the gauge condition of the form $D_\mu B^{\mu\nu} = 0$), one can, in principle, obtain the classical solution for $B^a_{\mu\nu}$. We, however, are explicitly interested in the spacetime metric solution which, as mentioned, is independent of the solution form of $B^a_{\mu\nu}$ and partly because $B^a_{\mu\nu}$ appears classically as a Lagrange multiplier enforcing the constraint $F^a_{\mu\nu} = 0$. Now suppose we look for static, spherically-symmetric solutions to the classical field equations that are asymptotically flat. Then, first the metric can be written in the form
\begin{equation}
ds^2 = -B(r)dt^2 + A(r)dr^2 + r^2d\Omega^2_2
\end{equation}
with $d\Omega^2_2$ being the metric on the unit two-sphere. Next, for the matter sector, especially for the vector potential solution of the pure gauge form in eq.(5), in order to look for a spherically-symmetric solution we take the standard ansatz which is the same in form as the flat spacetime Wu-Yang monopole solution ansatz \cite{7} (with the non-abelian gauge group for the anti-symmetric tensor field being chosen to be SU(2))
\begin{align}
A^a_0(r) &= 0, \nonumber\\
A^a_i(r) &= -\epsilon_{iab}\frac{x^b}{g_cr^2}[1-u(r)].
\end{align} 
As is well-known, this solution ansatz is indeed spherically-symmetric in the sense that the effect of a spatial rotation, SO(3), can be compensated by a gauge transformation, SU(2). In the spherical-polar coordinates, this ansatz for the vector potential $A^a_\mu$ and the non-vanishing components of the corresponding field strength $F^a_{\mu\nu}$ are given by
\begin{equation}
A^a_0 = A^a_r = 0, \ A^a_\theta = -\frac{1}{g_c}[1-u(r)]\hat{\phi}^a,\ A^a_\phi = \frac{1}{g_c}[1-u(r)]\sin\theta\hat{\theta}^a
\end{equation}
and
\begin{equation*}
F^a_{r\theta} = \frac{u'(r)}{g_c}\hat{\phi}^a, \ F^a_{r\phi} = -\frac{u'(r)}{g_c}\sin\theta\hat{\theta}^a,\ F^a_{\theta\phi} = \frac{[u^2(r)-1]}{g^2_c}\sin\theta\hat{r}^a
\end{equation*}
where prime denotes the derivative with respect to $r$ and 
\begin{align*}
\hat{r}^a &= (\sin\theta\cos\phi,\sin\theta\sin\phi,\cos\phi),\\
\hat{\phi}^a &= (\cos\theta\cos\phi,\cos\theta\sin\phi,-\sin\theta),\\
\hat{\phi}^a &= (-\sin\phi,\cos\phi,0).
\end{align*}
Here, it is interesting to note that the case $u(r) = 0$ corresponds to the exact albeit singular monopole solution of Wu-Yang type with non-vanishing $A^a_\mu$ and $F^a_{\mu\nu}$;
the case $u(r) = +1$ corresponds to the "trivial" vacuum solution with vanishing $A^a_\mu$ and $F^a_{\mu\nu}$; and finally the case $u(r) = -1$ corresponds to a "non-trivial" vacuum solution with vanishing $F^a_{\mu\nu}$ but non-vanishing $A^a_\mu$. Therefore, since we are looking for a non-trivial pure gauge solution satisfying $F^a_{\mu\nu}$ but non-vanishing $A^a_\mu$. Therefore, since we are looking for a non-trivial pure gauge solution satisfying $F^a_{\mu\nu}$ = 0, we should take the last case with $u(r) = -1$. Further one can easily check that this non-trivial vacuum gauge solution $A^a_0 = 0,\ A^a_i = -\epsilon_{iab}(2x^b/g_cr^2)$ does satisfy the necessary condition that it must satisfy, $D^\mu(\sqrt{g}A^a_\mu)$ = 0 in eq.(9). Now that we have established the spherically-symmetric vector potential solution to field equations in curved spacetime. As mentioned earlier, then, our next job is to substitute this non-trivial vector potential solution into the Einstein field equation in (10) to solve for the spacetime metric solution. The resulting Einstein equation now reads
\begin{align*}
&R_{\mu\nu} = 2\pi(A^a_\mu A^a_\nu),\\
&T_{\mu\nu} = \frac{1}{4}[A^a_\mu A^a_\nu - \frac{1}{2}g_{\mu\nu}(A^a_\alpha A^{a\alpha})]
\end{align*}
with
\begin{equation*}
A^a_0=A^a_r=0,\quad A^a_\theta=-\frac{2}{g_c}\hat{\phi}^a, \quad A^a_\phi=\frac{2}{g_c}\sin\theta\hat{\theta}^a.
\end{equation*}
Note that in terms of the isotropic metric given in eq.(11) only two components of the Einstein equations out of the three are truly independent because the third component is satisfied automatically due to the energy-momentum conservation, $T^{\mu\nu}_{;\mu}$ = 0. Thus we consider the following two independent combinations convenient for solving the Einstein equations,
\begin{align}
&\frac{1}{AB}(AR_{tt} + BR_{rr}) = 8\pi[-T^t_t + T^r_r], \nonumber\\
&\frac{1}{2}\left(\frac{1}{B}R_{tt} + \frac{1}{A}R_{rr}\right) + \frac{1}{r^2}R_{\theta\theta} = 8\pi[-T^t_t].
\end{align}\\\\
The first combination yields $B(r) = A^{-1}(r)$ where we imposed the asymptotic flatness condition, $A(r)\rightarrow1,\ B(r)\rightarrow1$ as $r\rightarrow\infty$. On the other hand the second combination gives
\begin{equation*}
A(r) = \left[1-\frac{2M(r)}{r}\right]^{-1},
\end{equation*}
where $M(r)$ is to be determined from $\frac{dM(r)}{dr} = 4\pi r^2\rho_m(r) = (4\pi/g^2_c)$ with $\rho_m(r) = [-T^t_t]$. Namely, $M(r) = M + 4\pi r/g^2_c$ with the integration constant $M$ being identified with the total mass-energy of the system defined at the spatial infinity, $i^0$, namely the "ADM mass". Finally, the classical vector potential and the metric solution are given by
\begin{align}
&A=A_\mu dx^\mu = \frac{1}{g_c}\left[-2\tau_\phi d\theta + 2\sin\theta\tau_\theta d\phi\right],\\
&ds^2 = -\left[\left(1-\frac{8\pi}{g^2_c}\right) - \frac{2M}{r}\right]dt^2 + \left[\left(1-\frac{8\pi}{g^2_c}\right)-\frac{2M}{r}\right]^{-1}dr^2 + r^2d\Omega^2_2 \nonumber
\end{align}
where $\tau_r \equiv \hat{r}^a(\frac{\sigma^a}{2}),\ \tau_{\theta} = \hat{\theta}^a(\frac{\sigma^a}{2}),\ \tau_\theta = \hat{\phi}^a(\frac{\sigma^2}{2})$ with $\sigma^a$ being the Pauli spin matrices. Note that there is also a trivial vacuum solution with corresponding gauge potential and metric being given by $A_\mu = 0$ (or $u(r) = +1$) and the usual Schwarzschild solution respectively, Here it is interesting to recognize that although the two gauge potential solutions, trivial vacuum $A_\mu = 0$ and the nontrivial vacuum gauge $A_\mu = (-i/g_c)(\partial_\mu U)U^{-1}$, are related by a gauge transformation and hence produce the same field strength tensor $F_{\mu\nu} = 0$, the spacetime metrics generated by each of the two gauge choices above are not related by any coordinate transformation and thus produce distinct curvatures. This can be easily seen by evaluation the curvature invariant $I = R_{abcd}R^{abcd}$ with $a,\ b,\ c,\ d$ being indices associated with an orthonormal basis. In the same Schwarzschild coordinates, the curvature invariant of the usual Schwarzschild solution is given by $I = \frac{48M^2}{r^6}$ whereas that of the metric solution in eq.(15) turns out to be $I =16[2+(1+\frac{4\pi r}{g^2_cM})^2]\frac{M^2}{r^6}$. Now, we would like to examine the nature of the spacetime described by our metric solution in eq.(15). To do so we consider three cases in the following section.

\begin{figure}[h!]
\includegraphics[width=11cm]{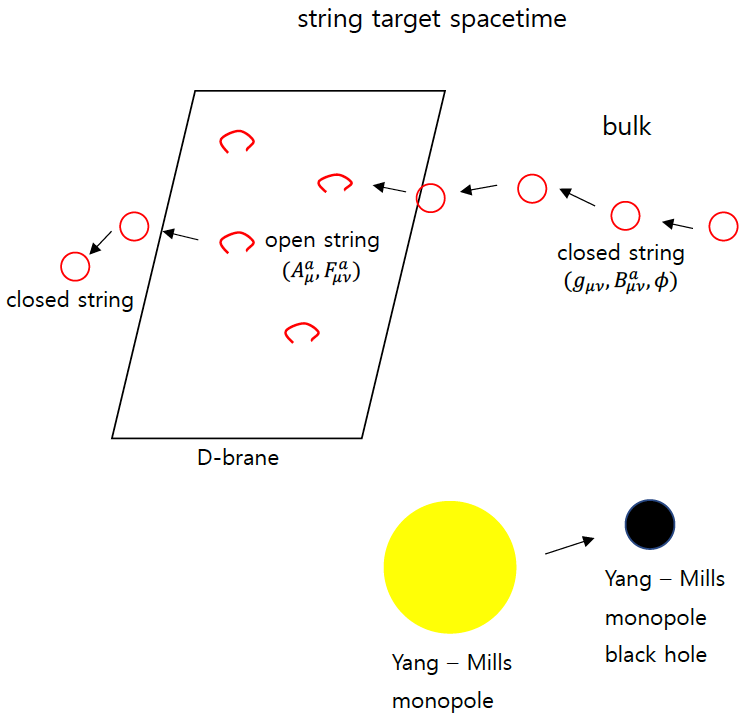}
\caption{Yang-Mills monopole solution and its evolution into a black hole in superstring theory}
\label{cartoon}
\end{figure}

\section{Yang-Mills monopole black hole solution in Superstring theory}
Upon finalizing the classical solution in the previous section, we now consider various cases of the solution categorized by different choices of parameters involved in this system.
\begin{itemize}
\item[(i)] In the weak coupling limit ($g_c \ll \sqrt{2\pi}$);
\begin{equation*}
ds^2 = \left[\left(\frac{8\pi}{g^2_c}-1\right) + \frac{2M}{r}\right]dt^2 - \left[\left(\frac{8\pi}{g^2_c}-1\right) + \frac{2M}{r}\right]^{-1}dr^2 + r^2d\Omega^2_2.
\end{equation*}
This metric represents a sapacetime in which $r$ is timelike and $t$ is spacelike. Thus the metric has an explicit time-dependence. The curvature singularity at $r = 0$ is timelike and the future of any Cauchy surface contains a naked singularity which is visible from the future null infinity $I^+$. Namely no event horizon arises and thus it exhibits an example of the violation of cosmic censorship hypothesis.
\item[(ii)] For the coupling constant $g_c = 2\sqrt{2\pi}$;
\begin{equation*}
ds^2 = \frac{2M}{r}dt^2 - \frac{r}{2M}dr^2 + r^2d\Omega^2_2.
\end{equation*}
Again this metric represents a spacetime in which $r$ is timelike and $t$ is spacelike. Also $r = 0$ is a naked singularity with no event horizon whatsoever around it and hence leads to the violation of the cosmic censorship hypothesis.
\item[(iii)] In the strong coupling limit ($g_c \gg 2\sqrt{2\pi}$);
\begin{equation*}
ds^2 = -\left[\left(1 - \frac{8\pi}{g^2_c}\right) - \frac{2M}{r}\right]dt^2 + \left[\left(1 - \frac{8\pi}{g^2_c}\right) - \frac{2M}{r}\right]^{-1}dr^2 + r^2d\Omega^2_2.
\end{equation*}
\end{itemize}
This metric describes a black hole spacetime with an event horizon placed at $r = 2M(1 - \frac{8\pi}{g^2_c})^{-1}$ which encloses a spacelike curvature singularity at $r = 0$. Since this metric is characterized by two parameters, $M$ and the non-abelian gauge coupling constant $g_c$, the black hole has a non-abelian hair. This black hole spacetime is, as emphasized, not merely a coordinate transformation of the usual Schwarzschild black hole but they have analogous global structures and thermodynamic properties. For instance, this black hole has Hawking temperature and entropy of $T_H = (1 - \frac{8\pi}{g^2_c})^2/8\pi M$ and $S = 4\pi M^2(1 - \frac{8\pi}{g^2_c})^{-2}$ respectively.\\
Now we conclude with few observations. Firstly, the "non-abelian" hair of the black hole solution in the strong coupling limit possesses an exotic property. Unlike the abelian gauge charge in the familiar Einstein-Maxwell theory, the non-abelian gauge coupling parameter $g_c$ that characterizes the black hole solution above is not measurable as surface integrals at spatial infinity. This is because the metric solution is coupled to the vacuum gauge solution $F_{\mu\nu}$ = 0 in the present theory. Secondly, the metric solutions for cases (i) and (ii) are shown to exhibit the violation of cosmic censorship hypothesis. They, in fact, provide non-trivial counter-examples to the hypothesis in the sense that both the physical and the mathematical versions of the hypothesis are violated. Namely, its classical metric solution turns out to violate the hypothesis while the present theory itself satisfies the dominant energy condition (i.e., the locally non-negative matter energy density), $T_{\mu\nu}n^\mu n^\nu = 1/g^2_cr^2 \geq 0$ (where $n^\mu$ is the timelike unit vector) on which the mathematical version of the hypothesis is based. Note that the cosmic censorship hypothesis is believed to hold in the classical theory of general relativity. And thus far there has been no known concrete example of the violation of the hypothesis with its origin being at the classical theory. "White holes", whose existence has been proposed to be possible, should not be regarded as a counter-example to the "classical" cosmic censorship hypothesis since they are objects that can be speculated to exist via the "time-reversal" of the classical black holes in the conventional definition or the quantum black holes that do evaporate in Hawking's option \cite{8}. In this sense, our classical metric solution in the present theory appears to be an interesting example that violates the hypothesis in the purely classical regime. It seems, however, fair to point out that the sort of the violation of the cosmic censorship hypothesis we found here is rather a peculiar consequence of the "exotic" classical metric solution that arises when a classical matter field theory is coupled to Einstein gravity than a phenomenologically realistic result arising from the gravitational collapse of some well-defined initial data.

 \section{Concluding Remarks} 
In our present revisit here, which is indeed a more enlarged/modern approach, basically the same set up should now be coined;
 the presence of both closed and open strings and coupling of their low-lying modes.
 To help one's understanding of the set up/situation that the present author speculates, if displayed by a motion picture, it goes like this: 
 at a certain point, the closed string starts floating/navigating the bulk, then bumps into and lands on a D-brane, slides around on it, then leaves/departs it back into the bulk.
 The closed string and the subsequent open string repeat this series of motions over and over again.
 This is precisely what the action and its associated classical solution to the Euler-Lagrange's equation of motion describe as a result of the coupling of low-lying modes of both the closed and open strings.
 
 To summarize, therefore, the result of our present work, namely, a manifest breakdown of features of classical gravitation and cosmology is really a direct consequence of the current full version of the superstring theory.
 Among others, it is interesting to note that unlike the purely classical treatment, that is, unlike in the conventional Einstein-Yang-Mills-dilation theory where the over-all environment is the Riemannian geometry/manifold, in the current full version of the superstring theory, now the over-all environment is Riemannian plus symplectic geometry/manifold as the closed string perturbations involve ($g_{\mu\nu}, B_{\mu\nu}, \phi$).
This realization may signal that the gravitation and cosmology sector of the full version of the superstring theory admits the built-in Mirror duality \cite{2, 3, 4, 5} which, again, is supposed to be a generic feature of a legitimate quantum theory of gravitation. Lastly, the author of the present work would like to leave one more comment. That is, aside from elevating his 1996 work into enlarged/modern version under the spirit of recent development of superstring theory, the addition of Yang-Mills monopole black hole content demanded highly involved contents of Yang-Mills monopole solution and self-gravitating black hole solution.

The useful lesson we learned from the present work is that unlike the classical theory of gravitation, the superstring theory which is the best candidate for theory of quantum gravity does not respect some of the cherished features of classical astrophysics and cosmology such as black hole no-hair theorem and cosmic censorship hypothesis. This is of course rather an anticipated results as any kind of viable theory of quantum gravity such as the superstring theory would depart from conventional ideas of classical gravitation.



\end{document}